\documentclass[twocolumn,showpacs,preprintnumbers,amsmath,amssymb,nofootinbib]{revtex4}
\input psfig.sty
\usepackage{graphicx}
\usepackage{dcolumn}
\usepackage{bm}
\usepackage{latexsym}
\usepackage{epsfig}

\newcommand{\be}{\begin{equation}}
\newcommand{\ee}{\end{equation}}
\begin{document}
\title{On the equivalence of $\Lambda(t)$ and gravitationally induced particle production cosmologies}

\author{L. L. Graef$^{1}$}\email{leilagraef@usp.br}
\author{F. E. M. Costa$^{2,3}$} \email{ernandesmc@usp.br}
\author{J. A. S. Lima$^{2}$}\email{limajas@astro.iag.usp.br}

\vskip 0.5cm
\affiliation{$^1$Instituto de F\'isica, Universidade de S\~ao Paulo, 05508-900, S\~ao Paulo, SP, Brazil}
\affiliation{$^2$Departamento de Astronomia, Universidade de S\~ao Paulo, 05508-900 S\~ao Paulo, SP, Brasil}
\affiliation{$^3$Universidade Federal Rural do Semi-\'Arido, 59900-000, Pau dos Ferros, RN, Brazil}

\date{\today}

\begin{abstract}
The correspondence between cosmological models powered by a decaying vacuum energy density and gravitationally induced particle production is investigated. Although being physically different in the physics behind them we show that both classes of cosmologies under certain conditions can exhibit the same dynamic and thermodynamic behavior. Our method is applied to obtain three specific models that may be described either as $\Lambda(t)CDM$ or gravitationally induced particle creation. In the point of view of particle production models, such cosmologies can be interpreted as a kind of one-component unification of the dark sector. By using current type Ia supernovae data, recent estimates of the cosmic microwave background shift parameter and baryon acoustic oscillations measurements we also perform a statistical analysis to test the observational viability within the two equivalent classes of models and we obtain the best-fit of the free parameters. By adopting the Akaike information criterion  we also determine the rank of the models considered here. Finally, the particle production cosmologies (and the associated decaying $\Lambda(t)$-models) are modeled in the framework of field theory by a phenomenological scalar field model.

\end{abstract}
\maketitle

\section{Introduction}

The current cosmic acceleration has been evidenced from distance measurements of type Ia supernovae data~\cite{data} and the simplest explanation is to admit the existence of a cosmological constant, $\Lambda$, which can be associated to the energy density stored in the true vacuum state of all existing fields in the Universe. From the observational point of view, it is well known that $\Lambda$ provides a very good description of the observed Universe. Despite its observational successes, it suffers at least from two problems. First, and possibly the most serious one is the cosmological constant problem (CCP). It refers to the fact that the cosmological upper bound ($\rho_{\Lambda} \lesssim 10^{-47}$ ${\rm{GeV}}^4$) differs from theoretical expectations for the vacuum energy ($\rho_{\Lambda} \sim 10^{71}$ ${\rm{GeV}}^4$) by approximately 120 orders of magnitude. The other is known as the coincidence problem and consists in understanding why $\rho_{\Lambda}$ is not only small, but also of the same order of magnitude of the energy density of cold dark matter (CDM) exactly today \cite{weinberg}.

A possible alternative to resolve the cosmological constant problems described above is to suppose that the vacuum energy is not a constant but decays into other cosmic components. Phenomenological models with variable cosmological term (decaying vacuum) have been proposed in literature as an attempt to alleviate the cosmological constant problem \cite{ozer,bert,CLW92,LM1,Wnew} and more recently the coincidence problem \cite{wang,alc05,costa77,costa81,costa12,leila12}. The usual treatment is to assume that $\Lambda(t) = 8\pi G \rho_v (t)$  behaves like a scalar field whose kinetic term is negligible while its potential energy is coupled to the other components of the universe thereby producing particles (the decay products) continuously and slowly. In these models the explanation for the present smallness of the vacuum energy density is that it has been decaying during the whole life of the universe, and, as such, the  vacuum energy density is small nowadays because the Universe is too old.  

The running of the vacuum energy density is generically expected from quantum field theory(QFT) in curved space-times \cite{new1,new2}. In this case, the effective QFT action implies that the variation of $\rho_{\Lambda}$ can be associated with the change of the space-time curvature, whose expression may depend on the specific gravity theory adopted. Many proposals along these lines, in which the presently observed value of $\Omega_{\Lambda}$ is a remnant from inflation, were discussed in the literature (for a recent review see \cite{new3}). In the Starobinsky model, for instance, the vacuum effective action of a massive scalar field is calculated by using a conformal representation of the field action \cite{new4} and leads (in the one-loop approximation) to a non-singular de Sitter stage, as generated in many phenomenological decaying vacuum cosmologies \cite{LM1,CTnew}. Cosmologies with extra dimensions also suggest that modifications of the Friedmann equation can mimick a time varying $\Lambda$(t)-term scaling as $H^{a}$, where $a$ is a constant index \cite{new5}. More recently, the cosmic variation of the vacuum energy density has also be justified based on the renormalization group (RG) approach \cite{new6}. Such attempts are collectively suggesting that dynamical $\Lambda$(t)-models provide an interesting possibility not only to accelerate the Universe but also to solve both the CC and coincidence problems.


Nevertheless, even in the context of general relativity, there are other possibilities to explain the present accelerating stage without dark energy thereby  evading the questions related to the cosmological constant problems \cite{Lima2004}. This happens, for instance,  when the matter content of the universe is subjected to some kind of dissipative process (a kind of  cosmic antifriction)  that can be expressed in the Einstein Field Equations (EFE) by the inclusion of an effective negative pressure \cite{dissip}. As a consequence,  a  late time accelerating stage appears naturally with the model providing a new alternative scenario to confront with the present day astronomical observations.  

Another interesting possibility is the phenomenon of gravitationally induced particle production at the expenses of a time varying gravitational field. From a microscopic viewpoint it has been justified after the pioneering works of Parker and collaborators \cite{new2}. As discussed by many authors, the positive and negative frequency of the fields in the Heisenberg picture become mixed during the universe expansion. As a result, the creation and annihilation operators at one time $t_{1}$ are linear combinations of those ones at an earlier time $t_{2}$, thereby resulting in particle production. Qualitatively, one may say that the time varying gravitational background works like a 'pump' supplying energy for the matter fields (see \cite{n2} for particle production in F(R) theories of gravity). Since the energy of the field is not conserved its action is explicitly time-dependent, with the quantization leading generically to particle production \cite{new2,n1,n3}.

Macroscopically, as originally discussed by Prigogine and coworkers \cite{Prigogine} based on non-equilibrium thermodynamics of open systems, this kind of process can also be described by a negative creation pressure (see also \cite{LCW,LG92}). In this case, by assuming that dark matter particles are produced by a time varying gravitational field, it is also possible to obtain a late time acceleration in a universe composed only by pressureless fluids, like baryons and cold dark matter \cite{LSS,Debnath11}. In the same vein, some authors also showed that the evolution of an arbitrary $\Lambda$CDM model can fully be mimicked by a baryonic fluid plus creation of CDM particles (CCDM model) both at background and perturbative levels \cite{limaccdm,Jesus10}. In the flat case, for instance, the CCDM model has also only one free parameter (like the standard $\Lambda$CDM) which describes the CDM particle production rate. Therefore, it is simple like the cosmic concordance model, evolves with the same dynamics, and, more important, it has only one component filling the dark sector (CDM) whose observational status is relatively higher than any kind of dark energy \cite{RCDM}.


Recently, Mimoso and Pav\'on \cite{MP13} also investigated the  thermodynamic behavior of two different classes of cosmologies: (i) a complete CCDM scenario \cite{LBC12}, and (ii) a complete decaying $\Lambda(t)$ scenario \cite{LBS13}. 
The quoted  authors concluded that these particular $\Lambda(t)$ and CCDM cosmologies are thermodynamically consistent even when  the horizon entropy  (during the extreme de Sitter phases)  are taken into account \cite{MP13}.     

In this paper we go one step further by investigating  whether there is a general  equivalence between $\Lambda(t)$CDM and CCDM cosmologies both from  a dynamic, as well as, from a thermodynamical viewpoint.  Our  basic interest here is to determine what are the general conditions under which both scenarios can provide the same cosmological description. As we shall see, there are general relations among the physical parameters involving the creation rate of CCDM cosmologies and the $\Lambda(t)$ model  which may guarantee, from the very beginning, the same macroscopic behavior, even considering that they are deeply different in the physics behind them. The equivalence in a perturbative level, i.e. by taking into account the evolution of the density fluctuations, will be investigated in a forthcoming communication (for a discussion involving only the CCDM approach see Ref. \cite{Jesus10}).

The manuscript is organized as follows. In section II we analyse the dynamic equivalence between the CCDM models and the decaying vacuum models. The thermodynamic equivalence is discussed in section III. In section IV we specialize our results by considering three distinct decaying vacuum models and the corresponding dynamics in CCDM models. In section V we compare CCDM models with observational results. In section VI we interpret this class of models in terms of the dynamics of an ordinary scalar field, and, finally,  in section VII, we summarize the basic results.

\section{Dynamic Equivalence}

The Einstein Field Equations relates the dynamic properties of a given spacetime with its total energy content (in our units $8\pi G \equiv c=1$)
\begin{equation} \label{efe}
G^{\mu \nu} = T^{\mu \nu},
\end{equation}
where $G^{\mu \nu}$ is the Einstein tensor and $T^{\mu \nu}$ is the total energy-momentum tensor of the cosmic fluid. 

In what follows, we will compare in detail  two different classes of models in the framework of a Friedmann-Lemaitre-Robertson-Walker space-time. 

Firstly, we will consider a generic decaying vacuum model whose thermodynamic behavior was discussed long ago by one of us \cite{Lima96}.  In this case, the EFE reduce to \cite{ozer,bert,CLW92}:
\begin{equation} \label{energia1}
{\rho} + \Lambda(t) = 3\frac{\dot{a}^2}{a^2} + 3\frac{k}{a^2}\;,
\end{equation}
\begin{equation}\label{pressure1}
p - \Lambda(t) = -2\frac{\ddot{a}}{a} - \frac{\dot{a}^2}{a^2} - \frac{k}{a^2}\;,
\end{equation}
where $\rho$ and $p$ are the energy density and the equilibrium thermostatic pressure of the usual cosmic fluid (baryons, radiation and dark matter) with $p = w \rho $, $a$ is the cosmic scale factor and $k$ is the parameter of curvature. For simplicity, henceforth it will be assumed that the decaying vacuum is coupled only with the dominant component.

The decaying vacuum causes a change in the number of particles of dark matter, so the equation describing particle concentration has a source term, i.e.,
\begin{equation}\label{conc}
N ^\alpha _{;\alpha} = \dot{n} + 3\frac{\dot{a}}{a} n = n \Gamma\;.
\end{equation}
Here, $\Gamma$ is the rate of change of the number of particles, $n=N/a^{3}$ is the particle number density and $N^{\alpha} = n u^{\alpha}$ is the particle flux.

By combining Eqs. (\ref{energia1}) and (\ref{pressure1}), or more directly, from the total energy conservation law one finds 
\begin{equation} \label{ec}
\dot{\rho} + 3\frac{\dot{a}}{a}(\rho + p) = - \dot{\rho}_\Lambda \;.
\end{equation}

Since the vacuum decay is the unique source of particle creation, we can write 
\begin{equation} \label{zeta}
\dot{\rho}_\Lambda = -\zeta n\Gamma,
\end{equation}
where $\zeta$ is a positive phenomenological parameter.

As remarked earlier, the second class of scenarios to be considered here are models with gravitationally induced particle production, sometimes named CCDM models \cite{LSS,limaccdm}. In this case, the Friedmann equations take the following form \cite{LCW,LG92,LSS,limaccdm}:
\begin{equation} \label{energia}
\tilde{\rho} = 3\frac{\dot{a}^2}{a^2} + 3\frac{k}{a^2}\;,
\end{equation}
\begin{equation}\label{pressure}
\tilde{p} + p_c = -2\frac{\ddot{a}}{a} - \frac{\dot{a}^2}{a^2} - \frac{k}{a^2}\;,
\end{equation}
where $p_c$ (creation pressure) is a non-equilibrium correction term describing the particle production. From now on, a tilde  denotes the fluid component quantities of the CCDM model in order to distinguish its values from their possible $\Lambda(t)$CDM counterparts. 

The particle number density in this case is described by the equation
\begin{equation}
\tilde{N} ^\alpha _{;\alpha} = \dot{\tilde{n}} +3 \frac{\dot{a}}{a} \tilde{n} = \tilde{n} \tilde{\Gamma}\;,
\end{equation}
where $\tilde{\Gamma}$ is the rate of particle creation induced by the varying gravitational field.

By combining Eqs. (\ref{energia}) and (\ref{pressure}) it is also possible to obtain the equation expressing the energy conservation law ($u_\mu T^{\mu \nu}_{;\nu} =0$)
\begin{equation} \label{conservacao}
{\dot{\tilde{\rho}}} + { 3H(\tilde{\rho} + \tilde{p} + p_c)} = 0\;.
\end{equation}
In general the creation pressure can be written as \cite{LG92}
\begin{equation}\label{pc1}
p_{c} = -\alpha \frac{\tilde{n} \tilde{\Gamma}}{3H}\;,
\end{equation}
where $\alpha$ is a positive phenomenological coefficient related to the creation process.

In order to obtain the dynamics of decaying vacuum models, we can combine Eqs. (\ref{energia1}) and (\ref{pressure1}), resulting in
\begin{equation}\label{evodv}
\frac{\ddot{a}}{a} + \Delta\frac{\dot{a}^2}{a^2} + \Delta \frac{k}{a^2} -\frac{(1+w) \Lambda(t)}{2} = 0\;,
\end{equation}
where $\Delta=(3w+1)/2$. Similarly, for models with matter creation we can combine Eq. (\ref{energia}) with (\ref{pressure}) to give
\begin{equation}\label{evoc}
\frac{\ddot{a}}{a} + \Delta \frac{\dot{a}^2}{a^2} + \Delta \frac{k}{a^2} + \frac{p_c}{2} = 0\;.
\end{equation}

Now, by comparing Eqs. (\ref{evodv}) and (\ref{evoc}) it is readily seen that the condition to a dynamic equivalence  is given by:
\begin{equation}\label{geral}
 p_c = - (1+w) \Lambda (t) \;.
\end{equation}
The above expression relates $\Lambda_0$ or general $\Lambda(t)$ cosmologies with the corresponding creation pressure of CCDM models. It is important to emphasize that the above condition is quite general and can be applied regardless of the phenomenological laws adopted to the decaying vacuum or to the creation rate describing the particle production. 

On the other hand, as discussed in Ref. \cite{LCW}, special attention has been paid to the simpler process termed ``adiabatic" particle production (see also \cite{LSS,Lima96}). It means that particles and entropy are produced in the space-time, but the specific entropy (per particle) remains constant $(\dot{\tilde{\sigma}} = 0)$. In this case, the constant $\alpha$ in equation (\ref{pc1}) is equal to $(\tilde{\rho} +\tilde{p})/\tilde{n}$, so that the creation pressure reads\footnote{ It should be noticed that fluids  endowed with ``adiabatic'' particle production satisfy the null energy condition (NEC) only if  ${\tilde{\Gamma}}/3H \leq 1$. All models that will be discussed in section IV satisfy such a condition.}
\begin{equation}\label{crea}
p_c = - \frac{(\tilde{\rho} + \tilde{p})\tilde{\Gamma}}{3H} = -\frac{(1+w) \tilde{\rho} \tilde{\Gamma}}{3H} \;.
\end{equation}
From Eqs. (\ref{geral}) and (\ref{crea}) we find
\begin{equation}\label{eqdvc}
{\Lambda (t)} = \frac{{\tilde{\rho}}{\tilde \Gamma}}{3H}\;.
\end{equation}
Note that above identification holds regardless of the curvature of the Universe. By assuming a spatially flat geometry, we have that $\tilde{\rho} = 3{H^2}$. Thus
\begin{equation}\label{eqvc}
\frac{\Lambda}{H^{2}} = \frac{\tilde{\Gamma}}{H} \;,
\end{equation}
which corresponds to a special case of Eq. (\ref{eqdvc}). In particular if $\Gamma<<H$ we find that $\Lambda<<H^{2}$, and, as such, both processes are negligible in this limit, as should be expected. 

\section{Thermodynamic correspondence}

Given that the dynamic equivalence is guaranteed by condition (\ref{geral}), or equivalently, by  (\ref{eqdvc}) in the case of adiabatic particle creation, let us now examine  the possibility of a complete thermodynamic equivalence.  The thermodynamic behavior of  $\Lambda(t)$ and particle production cosmologies were  discussed long ago (see  Refs. \cite{LG92,Lima96}), however, in a quite separated way, i.e. with no attempt to determine their possible equivalence.

In order to obtain the thermodynamic description of decaying vacuum-$\Lambda(t)$ models one needs to obtain the evolution equations of the specific entropy ($\sigma=S/N$) and temperature ($T$) of the created component. In this context, the vacuum  works like a second component transferring energy continuously to the matter component with the whole process constrained by the second law of thermodynamics. Following Lima \cite{Lima96}, we also assume that its chemical potential is null ($\mu_v=0$) so that the vacuum is a kind of  condensate carrying no entropy. Actually, for a null chemical potential,  the vacuum equation of state  ($p_v = -\rho_v$) implies that $\sigma_v=0$. Under such conditions,  the time-comoving derivative of the entropy flux, which is given by $S^{\alpha}=n\sigma u^{\alpha}$,  combined with  (\ref{conc}), (\ref{zeta}) and the Friedmann equations implies that
\begin{equation}\label{variacaos}
\dot \sigma + \sigma \Gamma = \frac{\Gamma}{T} (\zeta - \mu)\;,
\end{equation}
where $\mu$ denotes the chemical potential of the created particles, while the temperature $T$ satisfies the following evolution law:
\begin{equation}\label{evoltemp1}
\frac{\dot T}{T} = \biggl({\partial p \over \partial
\rho}\biggr)_{_n}\frac {\dot n}{n} -  
\frac {\Gamma}{T\biggl({\partial \rho \over \partial T}\biggr)_{_n}}\biggl[T\biggl({\partial p \over \partial T}\biggr)_{_n} +
n\biggl({\partial \rho \over \partial n}\biggr)_{_T}  - \zeta n \biggr]\;,
\end{equation}
(see \cite{Lima96} for more details).

In the case of CCDM models the specific entropy is given by 
\begin{equation}
\label{dotsigma}
\dot{\tilde{\sigma}} + \tilde{\sigma} \tilde{\Gamma} = \frac {\tilde{\Gamma}}{\tilde{T}}(\alpha - \tilde{\mu})\;,
\end{equation}
and the temperature follows the same evolution law as in the previous case \cite{LG92}
\begin{equation}
\label{evoltemp2}
       {\dot {\tilde{T}} \over \tilde{T}} = \biggl({\partial \tilde{p} \over \partial
\tilde{\rho}}\biggr)_{_{\tilde{n}}}\frac {\dot {\tilde{n}}}{\tilde{n}} -  
\frac {\tilde{\Gamma}}{\tilde{T}\biggl({\partial \tilde{\rho} \over \partial \tilde{T}}\biggr)_{_{\tilde{n}}}}\biggl[{\tilde{T}}\biggl({\partial \tilde{p} \over \partial \tilde{T}}\biggr)_{_{\tilde{n}}} +
\tilde{n}\biggl({\partial \tilde{\rho} \over \partial \tilde{n}}\biggr)_{_{{\tilde{T}}}}  - \alpha \tilde{n}\biggr]\;.
\end{equation}
Comparing Eqs. [(18)-(21)] we note that when $\alpha = \zeta$ the two pictures are thermodynamically equivalent. In addition, from equations (\ref{zeta}) and (\ref{pc1}) we also see that such an equality also implies $\dot{\rho}_{\Lambda} = 3Hp_c$, as should be expected due to the dynamic equivalence [compare Eqs. (\ref{ec}) and (\ref{conservacao})].

Now by considering that the particle creation process in both pictures is ``adiabatic", some equilibrium relations need to be preserved. In this case the second terms on right-hand side of Eqs. (\ref{evoltemp1}) and (\ref{evoltemp2}), which correspond to the non-equilibrium contributions, must be identically zero. In this case, it is possible to show that
\begin{equation}
\alpha = \zeta = \frac{\rho +p}{n}.
\end{equation}
Physically, this relation amounts to saying that $(\dot{\sigma} = 0)$. Hence, the equilibrium relations are preserved only if the specific entropy per particle of the created particles is constant. This means that
\begin{equation}\label{entro}
\frac{\dot{S}}{S} = \frac{\dot{N}}{N} = \Gamma \;,
\end{equation}
an expression valid for both pictures.

\section{Unifying the Dark Sector}

As seen previously, in vacuum decay models we must consider at least two main components, $\Lambda$ term and dark matter, in which vacuum is decaying. Now, we will interpret the standard model and some $\Lambda(t)$ models that have been discussed in literature in terms of matter creation models. We will restrict ourselves to late time behavior, and, as such, we take $w=0$.

Firstly, we rewrite Eq. (\ref{conservacao}) in terms of the interaction rate, i.e.,
\begin{equation} \label{conservacao2}
\dot{\tilde{\rho}} + 3H(\tilde{\rho} - \tilde{\Gamma} H) = 0\;.
\end{equation}
Generically, for a given $\Lambda(t)$ model, the corresponding matter creation model is obtained by combining Eq. (\ref{eqvc}) with (\ref{conservacao2}) and performing the integration.

\subsection{Case 1: Standard Cosmic Concordance Model}

Let us first consider the particular case $\Lambda(t)=cte=\lambda$, that corresponds to the standard $\Lambda$CDM model. In this case we have that
\begin{equation}\label{tlcdm}
\tilde{\Gamma} = \frac{\lambda}{H}\;,
\end{equation} 
where $\lambda$ is the cosmological constant of the $\Lambda$CDM model. Now, inserting above expression into Eq. (\ref{conservacao2}) and performing the integration, one finds
\begin{equation}\label{rho(a)I}
\tilde{\rho} = \lambda + \tilde{\rho}_{m1,0} a^{-3},
\end{equation}
where $\tilde{\rho}_{m1,0}$ is a constant with dimension of energy density that must quantify the current amount of matter that is clustering.
We can substitute the above relation into Eq. (\ref{energia}) in order to obtain an expression for $H$ as a function of redshift $(z)$, i.e.,
\begin{equation}\label{H(a)Ib}
{H} = H_0 \left[1 - \tilde{\Omega}_{m1} + \tilde{\Omega}_{m1}(1+z)^{3}\right]^{1/2},
\end{equation} 
where $\tilde{\Omega}_{m1} = \tilde{\rho}_{m1,0}/3H_0{^2}$. The above equation describes the dynamics of a CCDM scenario (CCDM1 in the present  notation), that behaves like the $\Lambda$CDM model, a result  previously derived by a different method \cite{limaccdm}. Naturally, as discussed in the first sections, due to the thermodynamic and dynamic equivalence it is rather difficult  to distinguish  observationally  between CCDM1 and $\Lambda$CDM model both at background and perturbative levels \cite{Jesus10}. However, from the theoretical viewpoint, they are quite distinct. In the $\Lambda$CDM model there are two main cosmic components that evolve independently of each other  thereby requiring a fine tuning. In the corresponding CCDM model, in turn, there is only one component, and so there are no problems of adjusting (for more details see, e.g., Refs.\cite{limaccdm,Jesus10,MP13,LBC12}).

\subsection{Case 2: $\Lambda = \gamma H$}

This simple phenomenological decaying vacuum law was proposed in Ref. \cite{saulo}, and, recently, discussed in a more general context \cite{1}. As remarked in the introduction, it can also be motivated e.g. as an intriguing and testable option for describing the present accelerating Universe, as well as a minimally modified Friedmann equation emerging from infinite-volume extra dimensions \cite{new5}. 
Note that $\gamma$ is a dimensional constant ($Dim\,{\gamma} \equiv Dim\, [H]$). In this case we find from Eq. (\ref{eqvc}) that the creation rate of particles is a constant, i.e.,
\begin{equation}\label{saulo}
\tilde{\Gamma} = \gamma \;.
\end{equation}
Consequently Eq. (\ref{conservacao2}) can integrated to give
\begin{equation}\label{rho(a)I}
\tilde{\rho} =  \frac{\gamma^{2}}{3} \left [ 1 + \left(\frac{C}{a}\right)^{3/2}\right]^{2},
\end{equation}
where $C$ is an integration constant. Now the Hubble parameter can be written as
\begin{equation}\label{hdez}
H = H_0 \left[1 - \tilde{\Omega}_{m2} + \tilde{\Omega}_{m2}(1+z)^{3/2}\right]\;,
\end{equation}
where $\tilde{\Omega}_{m2} = 1 - \gamma H_0 /3H_0^{2}$. This parameter quantifies the amount of matter that is clustering.

According to the second law of thermodynamics and Eq. (\ref{entro}) we see that $\Gamma \ge 0$. As $\Gamma = \gamma$, it implies that $\gamma > 0$.

\subsection{Case 3: $\Lambda = c + \beta H^{2}$}

\begin{figure*}[t]
\centerline{\psfig{figure=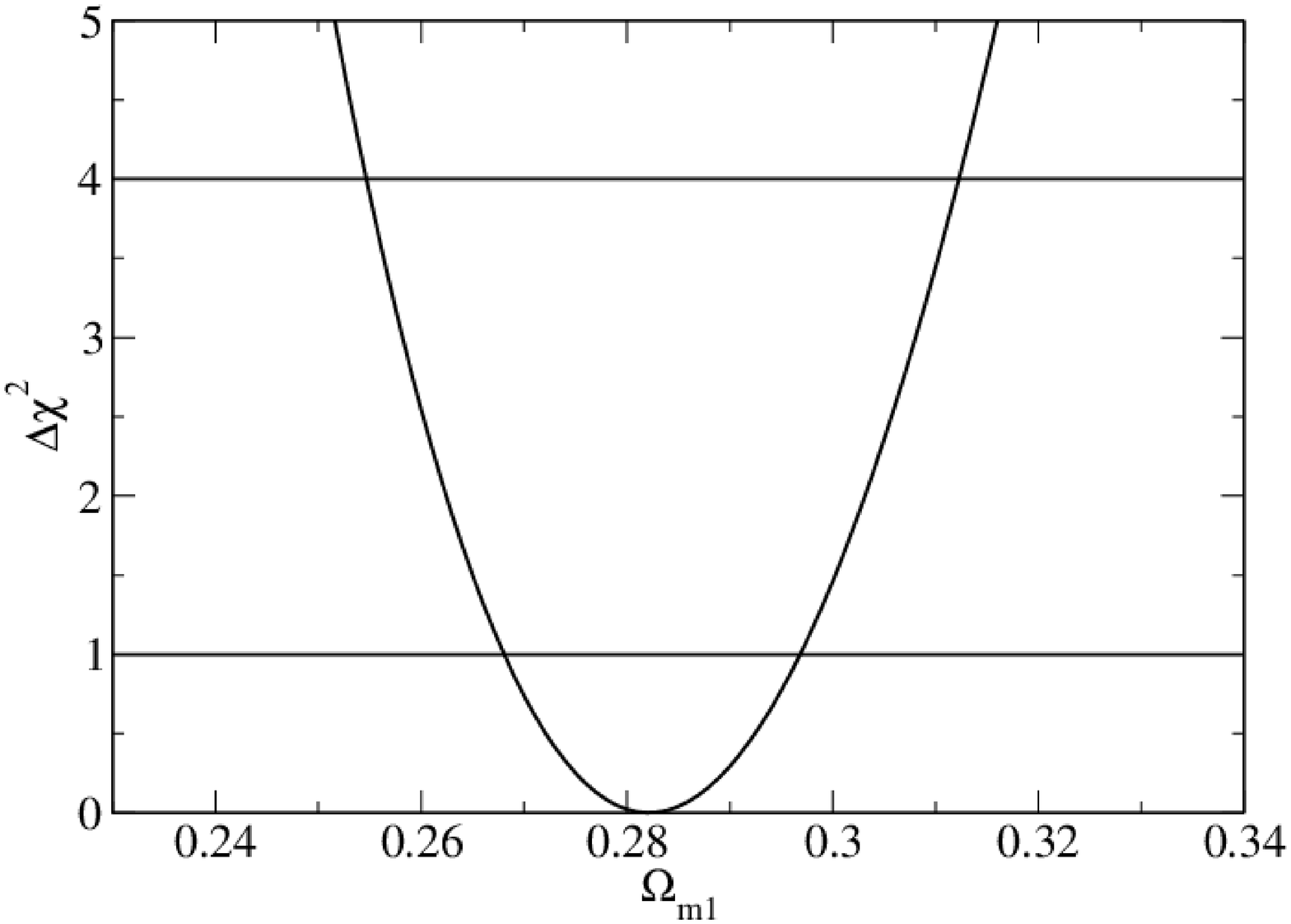,width=3.0truein,height=3.5truein,angle=-90}
\psfig{figure=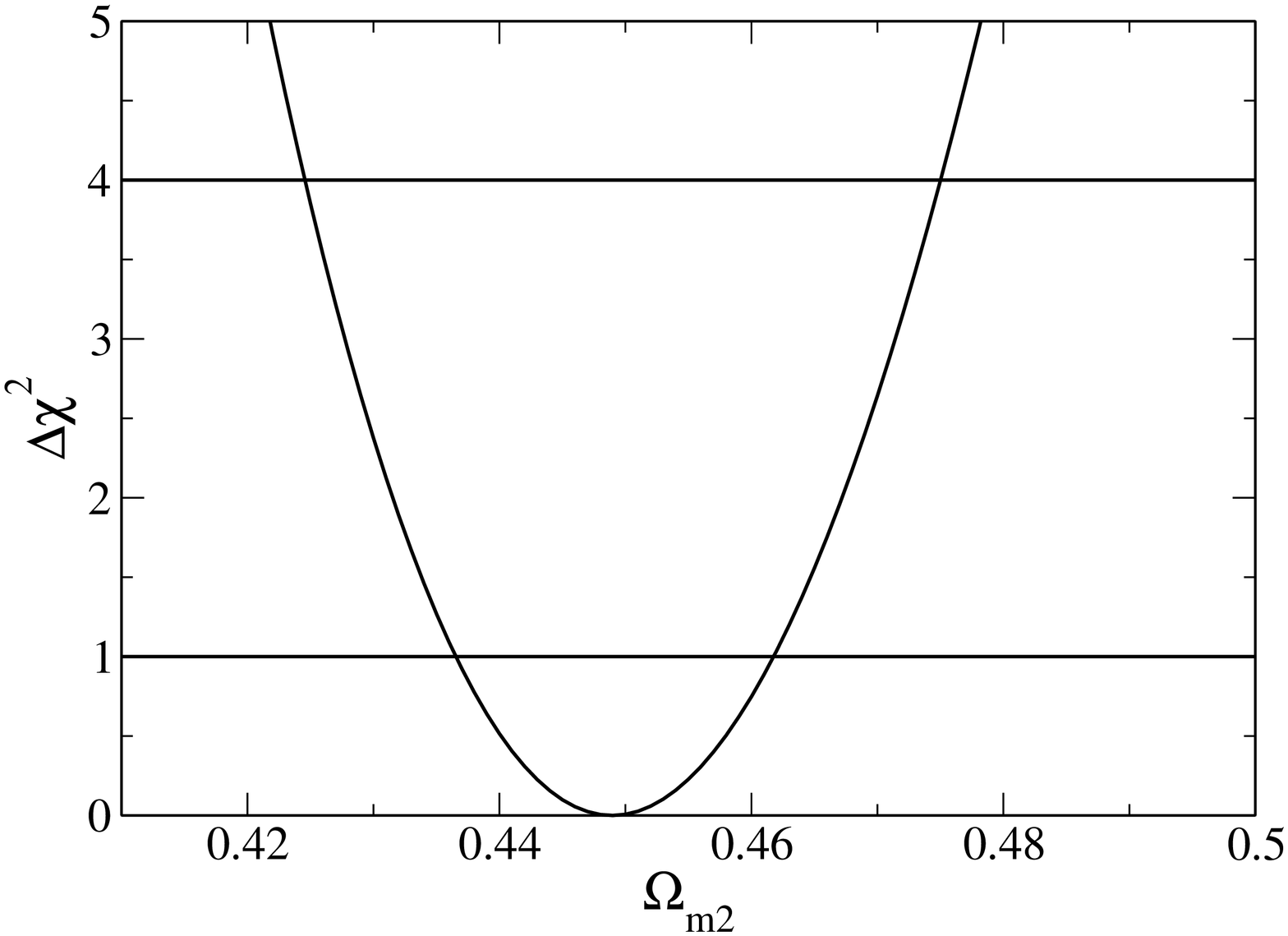,width=3.0truein,height=3.5truein,angle=-90}}
\caption{The variance $\Delta \chi^2$ as a function of the parameters $\tilde{\Omega}_{m1}\equiv \Omega_{m1}$ (left panel) and $\tilde{\Omega}_{m2} \equiv \Omega_{m2}$ (right panel). From this analysis, we find $\tilde{\Omega}_{m1} = 0.282^{+0.014}_{-0.014}$ and $\tilde{\Omega}_{m2} = 0.449^{+0.013}_{-0.013}$ at 1 $\sigma$ confidence level.}
\label{fig:qzw}
\end{figure*}

The $\beta H^{2}$ law was first phenomenologically proposed by Carvalho et al. in Ref.\cite{CLW92} and has also been extensively studied in the literature \cite{LM1,Wnew}.  Note that the late time evolution of this model is like the $\Lambda$CDM model at late times, however, the $H^{2}$ provides some dynamics even today thereby contributing to the constraints on the $\beta$ parameter\cite{LBS13,1}. As it appears, the model can also justified from first principles based on the renormalization group approach as describing the low energy physical running of $\rho_{\Lambda}(H)$.  In this context, the RG equation describing the vacuum energy density can be written as a series expansion in terms of $H$ \cite{shapiro,new6,1}:

\begin{equation}
\frac{d\rho_{\Lambda}}{d ln H^{2}}=\frac{1}{(4\pi)^{2}}\sum_{i} \left[a_{i}M_{i}^{2}H^{2} +b_{i}H^{4} + c_{i}\frac{H^{6}}{M_{i}^{2}} +... \right],
\end{equation}
where the sum over masses has been calculated in the one-loop approximation. The above expansion implies that the running proportional to $H^{2}$ (plus the additional constant) assumed here is quite suitable for the late time universe but not for the very early stages(for more details see \cite{1,new3} and Refs. therein). 

In this case Eq. (\ref{eqvc}) gives
\begin{equation}\label{shap}
\tilde{\Gamma} = \frac{c + \beta H^{2}}{H}\;.
\end{equation}
Inserting the above expression into Eq. (\ref{conservacao2}) and performing the integration, one finds
\begin{equation}\label{rho(a)I}
\tilde{\rho} = \frac{c}{1- \beta/3} + \tilde{\rho}_{m3,0}a^{-3 + \beta}.
\end{equation}

As in case I, we combine above relation with Eq. (\ref{energia}), so that 
\begin{equation}\label{H(a)Ib3}
{H} = H_0 \left[1 - \tilde{\Omega}_{m3} + \tilde{\Omega}_{m3}(1+z)^{3-\beta} \right]^{1/2}\;,
\end{equation}
where $\tilde{\Omega}_{m3} = \tilde{\rho}_{m3,0}/3H_0{^2}$.

At this point, it is also interesting to know the present value of the particle 
creation rate for the models discussed here (see Eq.(16)). By assuming that the CDM particles are neutralinos with mass $m \sim 100 Gev$, and that  $\Omega_{\Lambda_0} \sim 0.7$, 
$H_0 \sim  74 km.Mpc^{-1}.s^{-1}$ as suggested by the latest measurements \cite{Riess}, it is easy to check that the present creation rate is [$\Gamma n]_{today}\sim {10^{-11}}.cm^{-3}.yr^{-1}$. Such a rate has not appreciably been changed in the last few billion years ($z < 1$) when the Universe started to accelerate powered (in our description) by the particle production process \cite{limaccdm}.



To obtain some thermodynamic constraints on the parameter $\beta$ of the decaying vacuum relation $\Lambda = c + \beta H^{2}$, let us combine this result with Eqs. (\ref{energia1}) and (\ref{ec}), resulting in
\begin{equation}
\rho_{m} = \rho_{m,0}a^{-3+\beta}.
\end{equation}
As $\rho_{m} = m n$ and we are assuming that $m = constant$, we have that $n = n_{0}a^{-3+\beta}$. Replacing $n$ into Eq. (\ref{conc}) and combining the result with Eq. (\ref{entro}) it is possible to show that
\begin{equation}
\dot{S} = S_0 \beta a^{\beta} H.
\end{equation}
As $a^{\beta} > 0$ and the universe is expanding ($H > 0$), the second law of thermodynamics, $(\dot{S} \ge 0)$, implies that $\beta \ge 0$, neglecting the vacuum entropy.

\section{Observational comparison}

In order to constraint the parameters of the matter creation model considered here we make use of different data sets. The primary data set used in this analysis comprises recent SNe Ia compilation, the so-called Union 2.1 sample, compiled in ref.~\cite{union21} which includes 580 data points after selection cuts. 

Additionally, we also use measurements derived from the product of the CMB acoustic scale, and from the ratio of the sound horizon scale at the drag epoch to the BAO dilation scale.

The best fit to the set of parameters $s$ is found from the minimization of the function $\chi^2_{\rm{T}} = \chi^2_{\rm{SNe}} + \chi^2_{\rm{CMB/BAO}}$, where $\chi^2_{\rm{SNe}}$ and $\chi^2_{\rm{CMB/BAO}}$ correspond to the SNe Ia and CMB/BAO $\chi^2$ functions, respectively.

\begin{figure}[th]
	\centerline{\psfig{figure=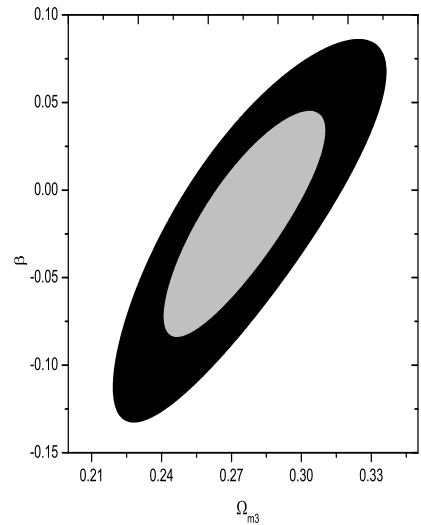,width=3.2truein,height=3.2truein,angle=0}}
\caption{The results of our statistical analysis in the plane $\beta - \tilde{\Omega}_{m3}$ for CCDM3. Constraints from SNe Ia and CMB/BAO ratio data are shown at $1\sigma$ and $2 \sigma$ confidence levels.}
\end{figure}

\subsection{Results}

The results of our statistical analyses are shown in Figs. 1 and 2. Figure 1 shows the variance 
$\Delta \chi^{2} = \chi^{2} - \chi^{2}_{min}$ at confidence regions (68.3\% CL and 95.4\%) for CCDM1 (left panel) and CCDM2 (right panel). For CCDM1 we find that the likelihood function peak is at $\tilde{\Omega}_{m1} = 0.282^{+0.014}_{-0.014}$, therefore, in excellent agreement with observations. While for CCDM2 the peak is at $\tilde{\Omega}_{m2} = 0.449^{+0.013}_{-0.013}$. This shows that a constant creation rate of particles [see Eq. (\ref{saulo})] during the cosmic evolution supplies a current value for $\tilde{\Omega}_{m}$ higher than the observed.



In Figure 2, we show the parameter space $\tilde{\Omega}_{m3} - \beta$ for CCDM3. By marginalizing on the nuisance parameter $h$ ($H_0=100hkm s^{-1}  Mpc^{-1}$) we find $\tilde{\Omega}_{m3} = 0.274^{+0.014}_{-0.014}$ and $\beta=-0.018^{+0.026}_{-0.027}$ at $68.3\%$ confidence level, with $\chi^{2}_{min}=563.53$ and $\nu=581$ degrees of freedom. While the reduced $\chi^{2}_{r} \equiv \chi^{2}_{min}/\nu = 0.97$, thereby showing that the model provides a very good fit to these data. 

\subsection{Viability of CCDM models}

Now, to compare the CCDM scenarios previously investigated, we use the Akaike information criterion (AIC), defined as
\begin{equation}\label{aic}
\rm{AIC} = -2\ln{\cal{L}} + 2k\; ,
\end{equation}
where ${\cal{L}}$ is the maximum likelihood and $k$ is the number of model parameters (see \cite{liddle, god, bie, davis} for reviews on the background for the use of this information criterion). As argued in Ref. \cite{liddle}, the AIC provide an interesting way to obtain a relative ranking of the observational viability of different candidate models. Thus, the important quantity in this analysis is the difference $\Delta {\rm{AIC}}_i =  {\rm{AIC}}_i - {\rm{AIC}}_{{\rm min}}$ calculated over the whole set of scenarios ($i = 1, ..., n$) with the best-fit model being the one that minimizes the {AIC} factor.

Table 1 shows a summary of the information criterion results for a SNe Ia sample and the CMB/BAO ratio data discussed above. As we can see the best-fit model is the CCDM1 model. Secondly we have the CCDM3 model, which is compatible with $\Lambda CDM$ model in 1 $\sigma$. 




\section{CCDM models and scalar field}

As seen previously, matter creation models can explain the cosmic acceleration without the introduction of a dark energy component. However, it is most desirable to represent them in a field theoretical language, i.e., in terms 
of the dynamics of an ordinary scalar field ($\phi$). 

In order to represent the matter creation models in terms of the dynamics of a scalar field, we replace $\tilde{\rho}$ and $\tilde{p}_{tot}=\tilde{p}+p_{c}$ in Eqs. (\ref{energia}) and (\ref{pressure}) by the
corresponding scalar field expressions
\begin{equation}
\tilde{\rho} \rightarrow \rho_{\phi} = \frac{\dot{\phi}^{2}}{2} + V(\phi), \;\;\;\;\;\;
\tilde{p}_{tot} \rightarrow p_{\phi} =\frac{\dot{\phi}^{2}}{2} - V(\phi) \;.
\end{equation}
Inserting the latter into the Friedmann's equations we can separate the scalar field
contributions and express them in terms of $H$ and $\dot{H}$, i.e.,
\begin{equation}
\dot{\phi}^{2} =-2\dot{H} \;,
\label{ff3}
\end{equation}
\begin{equation}
\label{Vz}
V={3H^{2}}\left( 1+\frac{\dot{H}}{3H^{2}}\right)=
{3H^{2}}\left( 1+\frac{aH^{'}}{3H}\right) \;,
\end{equation}
where $\dot{H}=aHH^{'}$ and prime denotes 
derivative with respect to the scale factor.
Now, considering that $dt=da/aH$, Eq. (\ref{ff3}) can be integrated
to give
\begin{equation}\label{ppz}
\phi=\int \left( -2{\dot{H}}\right)^{1/2} dt =
\int \left(-\frac{2H^{'}}{aH}\right)^{1/2}da\;.
\end{equation}

\begin{table}[]  
\begin{center}  
\caption{Summary of AIC results}
\begin{tabular}{rrllr}
\hline  \hline \\
\multicolumn{1}{c}{Model}&
\multicolumn{1}{c}{$k$}&
\multicolumn{1}{c}{Ranking}&
\multicolumn{1}{c}{$\Delta$AIC} &
\multicolumn{1}{c}{$\chi^2_{min}/\nu$} \\ \hline \\
CCDM1 & \quad  1 \quad \quad & \quad \quad 1 \quad \quad & \quad 0.00 & 0.97 \\
CCDM2 & \quad 1 \quad \quad &\quad \quad 3 \quad \quad & \quad 33.21 & 1.03 \\ 
CCDM3  & \quad 2  \quad \quad &\quad \quad 2 \quad \quad & \quad 1.86 & 0.97\\
\hline  \hline
\end{tabular} 
\end{center}
\end{table}

Having both the expressions for $V$ and $\phi$, it's easy to combine them in order to obtain the potential for the scalar field, $V(\phi)$, which represents the model in question. 

As an example, let's apply this procedure to obtain the scalar field description of the model CCDM3.

In this model the evolution of the Hubble function is given by
Eq. (\ref{H(a)Ib3}). Now, inserting Eq. (\ref{H(a)Ib3}) as well as its
derivative $(H^{'})$ into Eq. (\ref{ppz}) and integrating we obtain

\begin{equation}\label{phia3}
\phi(a) = \frac{1}{\sqrt{3-\beta}} \ln \left[\frac{\sqrt{A_3 a^{3-\beta} +1} -1}
{\sqrt{A_3 a^{3-\beta} +1}
+ 1}\right]\;,
\end{equation}
where $A_3 = (1 - \tilde{\Omega}_{m3})/\tilde{\Omega}_{m3}$. Now inserting Eq. (\ref{H(a)Ib3}) and it's derivative into Eq. (\ref{Vz}) we obtain the potential in terms of the scale factor, i.e.,
\begin{equation}\label{vdea3}
V(a) = 3{H_0^{2}} \left[1 -\tilde{\Omega}_{m3} +
\frac{\tilde{\Omega}_{m3}}{2}(1 +\beta /3)a^{\beta-3}\right]\;.
\end{equation}
Finally, comparing the equations above we find
\begin{equation}\label{vphif3}
V(\phi) =  D + E\cosh(\sqrt{3-\beta}\phi)\;,
\end{equation}
where $D = 3{H_0^{2}}({1-\tilde{\Omega}_{m3}})(3 - \beta/3)/4$ and $E = 3{H_0^{2}}({1-\tilde{\Omega}_{m3}})(3 + \beta/3)/4$. 

Exactly the  same procedure can be applied to the models CCDM1 and CCDM2, so we will present here only the final results. 
It's trivial to obtain that the CCDM1 model can be represented by a scalar field with the following potential
\begin{equation}\label{vphif}
V(\phi) =  B [3 +  \cosh(\sqrt{3}\phi)]\;,
\end{equation}
where $B = 3{H_0^{2}}({1-\tilde{\Omega}_{m1}})/4$. As one may check, this potential corresponds to the potential obtained for the model CCDM3 for $\beta = 0$.

While for the CCDM2 model the resulting scalar field potential has the following form
\begin{equation}\label{vphif2}
V(\phi) =  \frac{C}{8} \{2 + 6\cosh(\sqrt{3} \phi /2) + [\cosh(\sqrt{3} \phi /2) - 1]^{2} \}\;.
\end{equation}

So, as we can see, all these models can be represented by scalar fields with hyperbolic potentials.

\section{Conclusions}

In this work we have compared (in the context of the FLRW metric) the main dynamic and thermodynamic aspects of models with decaying vacuum ($\Lambda(t)$CDM) and gravitationally induced particle creation (CCDM). In particular,  we have established  under which conditions they exhibit the same dynamic and thermodynamic behavior. Using this equivalence we have reinterpreted the dark sector in terms of only one cosmic component (CDM). In order to exemplify the method  developed here we have found the CCDM models corresponding to three different classes of decaying vacuum models. All these equivalent CCDM cosmologies can be represented in terms of a  scalar  field description  whose potentials are given by hyperbolic functions.

By using current data, we have also performed a statistical analysis and showed that observationally the two pictures ($\Lambda(t)CDM$ and CCDM) are very similar. Based on the  Akaike information criterion (AIC) we have selected the best-fit and ranked the models considered here.

It is worth notice that  from thermodynamics the energy flow occurs from vacuum to the cold dark matter component. However, from an observational view point there is room for an interacting parameter which favors an energy flow in the opposite direction [cf. Fig. (2)]. Naturally, this result does not affect the  dynamic behavior of the models, but, in principle, at least for $\Lambda(t)$CDM models, it may suggest that the chemical potential of the created component may play a role not considered here \cite{PJ}.  

Finally, we stress  that in the  thermodynamic equivalence discussed here, the entropy associated to the possible existence of apparent horizons  were not taken into account. Its inclusion is somewhat natural when a de Sitter phase is present but it cannot be decided a priori since the solutions must be known (in this connection see Refs. \cite{MP13, LBC12,LBS13}).

\vspace{0.3cm}
{\bf Acknowledgments:} The authors are grateful to an anonymous referee for his/her comments and suggestions which improved the manuscript. L.G. is supported by FAPESP under grants 2012/09380-8.  F.E.M.C. is supported by FAPESP under grants 2011/13018-0, and J.A.S.L. is partially supported by CNPq and FAPESP under grants 304792/2003-9 and 04/13668-0, respectively.


\end{document}